\documentclass[epj]{svjour}
\usepackage{graphics}
\begin{document}
\title{Mott-Hubbard phase transition in 2D electron liquid}
\author{Igor N. Karnaukhov, Kateryna Levchuk, and Igor N. Dubinskiy% etc
\thanks{\emph{Present address:} karnaui@yahoo.com}%
}                     % Do not remove
\offprints{Igor N. Karnaukhov}          % Insert a name or remove this line
\institute{G.V. Kurdyumov Institute for Metal Physics, 36 Vernadsky Boulevard, 03142 Kiev, Ukraine}
\date{Received: date / Revised version: date}
\abstract{
We study the behavior of fermion liquid defined on hexagonal and triangular lattices with short-range repulsion at half filling.
In strong coupling limit the Mott-Hubbard phase state is present, the main peculiarity of insulator state is a doubled cell of the lattices.  In the insulator state at half filling fermions with momenta $k$ and $k+\pi$ are coupled via the effective $\lambda$-field, the gap in the spectrum of quasi-particle excitations opens and the Mott phase transition is occured at a critical value of the one-site Hubbard repulsion~$U_c$. $U_c=3.904$ and $U_c=5.125$ are calculated values for  hexagonal and triangular lattices, respectively. Depending on the magnitude of the short-range repulsion, the gap in the spectrum and the energy of the ground state are calculated. The proposed approach is universal; it is implemented for an arbitrary dimension and symmetry of the lattice for fermions models with short-range repulsion.
\PACS{
      {PACS-71.10.Fd}{Lattice fermion models (Hubbard model, etc.)}   \and
      {PACS-71.27.+a}{Strongly correlated electron systems; heavy fermions}
     } % end of PACS codes
} %end of abstract
\maketitle
\section{Introduction}
\label{intro}

We analyzed the Mott-Hubbard phase transition of fermion liquid defined in the hexagonal and triangular lattices under the on-site Hubbard repulsion.
This problem is physically important since graphene has the honeycomb structure and interacting electron liquid on triangular lattice is a frustration system. The study of the Mott phase transition has been immensely useful in understanding of behavior of strongly correlated systems \cite {Arovas,Karnaukhov2020,Karnaukhov2021,Karnaukhov2022,Cyrot,Chen}.

Despite it is a strikingly rich history of investigation the physical nature or scenario of the Mott phase transition  remains controversial. Should be noted that the Mott phase transition in the hexagonal lattice with on-site Hubbard interaction is investigated intensively, by means of numerical calculations \cite{Assaad,Meng,Sorella1,Sorella2,Paiva,Herbut}. Unfortunately, these calculations cannot be concluded about the spinless analogue of this model with the Coulomb repulsion between spinless fermions located at neighboring lattice sites. Monte Carlo simulations different authors give result for critical value of the on-site Hubbard interaction at which the Mott-Hubbard phase transition occurs equal to about 4.  What kind of phase transition (first or second) has not been unambiguously established? Despite the great interest in graphene and materials based on it, the Mott phase transition in the model with short-range repulsion on honeycomb lattice had not been study in detail. We do not know answer on the simplest question. What value of the short-range Coulomb repulsion does the gap open at in the spectrum?  It is necessary to know, because the stability of the state of graphene (with the Dirac spectrum in semi-metal state) is also determined by the repulsion between fermions which located at the nearest-neighbors lattice sites. The assertion that the symmetry protects gapless phase state is not an argument to be taken seriously, because the insulator phase is realized due the short-range repulsion is taken into account.

The peculiarity of the Mott-Hubbard phase transition on a triangular lattice has not been studied well enough (see numerical calculations in \cite{Kokalj,Yoshioka}). The problem further complicated by the fact that in this case we are talking about a system in the presence of frustration. In recent papers we shown that the cell doubling at the phase transition is the key to understanding nature of the Mott phase transition in interacting electron liquid which is determined on different lattices and for different dimension of the system \cite{Karnaukhov2020,Karnaukhov2021,Karnaukhov2022}.

In this paper we answer one of the important question. What is a scenario of the Mott-Hubbard phase transition or how does the phase transition occur? At a point of the phase transition the gap opens  at half filling, this phase transition is similar to the Peierls transition. The effective $\lambda$-field is determined by an unknown phase. We found the phase, it equal to $\pi$, what corresponds the cell doubling. We are not discussing the effect of impurities on the stability of the phase state at this time.

\section{Model and method}
\label{sec:1}
The model Hamiltonian is determined on the honeycomb and triangular lattices, includes the on-site Hubbard interaction
\begin{eqnarray}
{\cal H}= -
\sum_{\sigma=\uparrow,\downarrow}\sum_{<i,j>} c^\dagger_{i,\sigma} c_{j,\sigma}+  U\sum_{j}n_{j,\uparrow}n_{j,\downarrow}-\mu \sum_{\sigma=\uparrow,\downarrow}\sum_j  n_{j,\sigma},
\label{eq1}
\end{eqnarray}
where $c^\dagger_{j,\sigma},c_{j,\sigma}$ are the fermion operators determined on a lattice site $j$ with the spin $\sigma=\uparrow,\downarrow$, $U$ is the  value of the on-site Hubbard interaction determined by the density operator $n_{j,\sigma}=c^\dagger_{j,\sigma}c_{j,\sigma}$, for the tetragonal lattice at half-filling the chemical potential $\mu$ is equal to zero, in the case of a triangular lattice it is nonzero in a gapless state.

Using the Hubbard-Stratonovich transformation  we redetermine the interaction term $-U\sum_j \chi^\dagger_j \chi_j$, here
$\chi_j=c^\dagger_{j\uparrow}c_{j,\downarrow}$  in the following form $\sum_j(\frac{\lambda^\dagger_j\lambda_j}{U}+ \lambda_l\chi^\dagger_j+\lambda^*_j\chi_j)$. In the form $\lambda_j=\lambda \exp(i\textbf{q}\cdot\textbf{j})$  the solution for $\lambda$-field does not change chiral symmetry of the model at $q_x=q_y=0$ and
$q_x=2\pi$, $q_y=0$ or $q_x=0$, $q_y=2\pi/\sqrt{3}$.
Numerical calculations of the Hamiltonian (\ref{eq1}), defined on hexagonal and triangular lattices, show that trivial solution of $q=0$ describes noninteracting electron liquid with shifted branches of quasi-particle excitations. Below we consider the states of electron liquid at $q=\pi$, that determine new phase state with a renormalized spectrum.

Taking into account the solution  for $q_x=2\pi$, $q_y=0$ or $q_x=0$, $q_y=2\pi/\sqrt{3}$, we consider the solution of the Hamiltonian (1), defined on the hexagonal lattice, as the low energy effective Hamiltonian ${\cal H}_{eff}$. The effective Hamiltonian is defined by the $\lambda$-field as
\begin{equation}
{\cal H}_{eff}(\textbf{k,q}) = \left(
\begin{array}{cccccc}
\varepsilon&- w(\textbf{k})& \lambda & 0 \\
-w^*(\textbf{k})& \varepsilon &0 &\lambda^*\\
\lambda^* & 0 & \varepsilon &-w(\textbf{k}+\textbf{q})\\
0&\lambda&- w^*(\textbf{k}+\textbf{q}) & \varepsilon
\end{array}
\right),
\label{eq2}
\end{equation}
where $w(\textbf{k})=1+2\cos(k_x/2)\exp(ik_y \sqrt {3}/2)$, $\textbf{k}$ is the wave vector and $w(k_x+2\pi,k_y)=w(k_x,k_y+2 \pi/\sqrt{3})$.

At half filling the chemical potential is equal to zero for arbitrary $\lambda$  and $\textbf{q}$, the spectrum of
the quasi-particle excitations $\pm\varepsilon_{1,2}(\textbf{k,q})$ is symmetric about zero energy.

$$\varepsilon^2_{1,2}(\textbf{k,q})=\frac{1}{2}\left(|w(\textbf{k})|^2+|w(\textbf{k+q})|^2+2| \lambda|^2 \pm\sqrt{(|w(\textbf{k})|^2-|w(\textbf{k+q})|^2)^2+
4|\lambda|^2|w(\textbf{k})+w(\textbf{k+q})|^2} \right),$$

only for separated values of the $\textbf{q}$-vector its chirality is conserved, here. The $ \lambda $-field removes the spin degeneracy of the electronic spectrum.

The action has the following form
\begin{eqnarray}
\frac{S}{2 \beta}=-\frac{{T}}{{2}}\sum_{\textbf{k}}\sum_n \sum_{\gamma=1,2} \ln [\omega^2_n+\varepsilon^2_\gamma(\textbf{k,q})]
%\nonumber\\&&
+\frac{|\lambda|^2}{U},
 \label{eq3}
\end{eqnarray}

where $\omega_n =T(2n+1)\pi$ are the Matsubara frequencies,  quasiparticle excitations $\pm\varepsilon_\gamma(\textbf{k,q})$ ($\gamma =1,2$)
determine the fermion states in the
$\lambda$-field with the effective Hamiltoninan (\ref{eq2}), two on the right part of (\ref{eq3}) takes into account cell doubling and on the left, the spin degree of freedom.
In the saddle point approximation $\lambda$ is
the solution of the following equation $\partial S/\partial \lambda =0$.
At half filling occupation an equation, which corresponds to the minimal action $S$, is reduced to  the following at $T=0$~K

\begin{equation}
\frac{2|\lambda|}{U}=- \frac{1}{2}\sum_{\gamma=1,2}\int d \textbf{k}
\frac {\partial \varepsilon_\gamma (\textbf{k,q})}{\partial |\lambda|}.
\label{eq4}
\end{equation}

Equation (\ref{eq4}) allows one to determine the effective constant $ \lambda $ for a given bare value of the Hubbard interaction $U$.\\

\section{Behavior of an electron liquid on a hexagonal lattice}
\label{sec:2}

At half-filling the spectrum of the quasi-particle excitations $\varepsilon^2_{1,2}$ is symmetric about zero energy for arbitrary $\lambda$ also, it has four different branches, as a result, the cell doubles in the insulator phase. The behavior of the low energy branches of the spectrum determine the phase state of the electron liquid. Such at $0<\lambda<1$ the spectrum is gapless, at $\lambda_s=1$ a gap opens, the value of the gap $\Delta$ is equal to $2(\lambda -1)$.

\begin{figure}[!ht]
\centering
\vspace{4cm}
\includegraphics{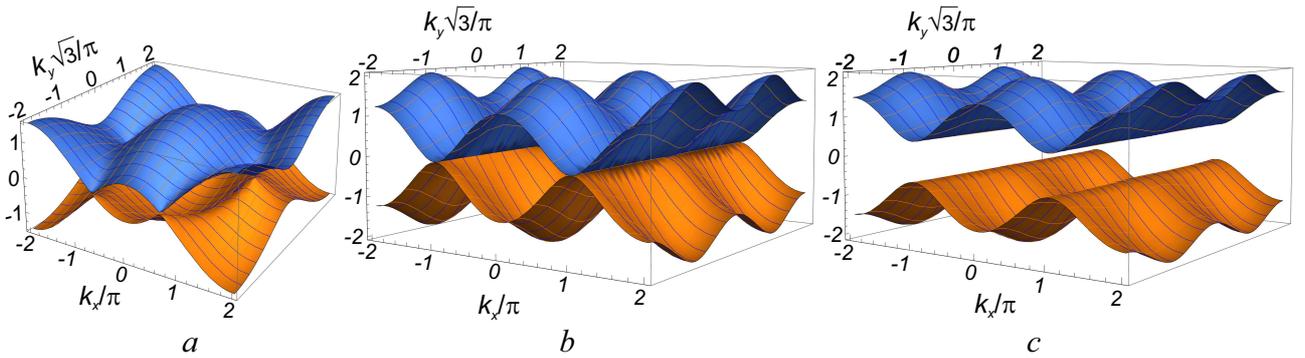}
%\vspace{4.25cm}
\caption{(Color online) Fermion spectrum is calculated on a hexagonal lattice at $U=0$ (gapless state) $a$), at $\lambda_s$=1 (state when the gap opens) $b$) and at $\lambda$=1.5 when gap is opened, here $\Delta$=1 $c$)
}
\label{fig:1}
\end{figure}

As is known, the electronic spectrum of graphene, without taking into account the interaction between fermions forms a spatial hexagon is bounded by the first Brillouin zone (Fig.~\ref{fig:1},~$a$). It is located relative to the $z$-axis at the Fermi energy level. Simultaneously, these surfaces represent the structure of the valence and conduction bands, which are in contact at six Dirac points (at the edges of the Brillouin zone). The valence band is completely filled with fermions and the conduction band is empty. Under the influence the $\lambda$-field the lattice cell doubles and  four different branches of quasi particle excitations are formed.
In the gapless phase with an increase in $\lambda$ from 0 to 1, the low-energy spectrum is converted. The distance between the three Dirac-Brilouin points in one line along the ordinate axis ($k_y$) for each of the pairs gradually decreases along the abscissa axis ($k_x$) and at $\lambda=1$ these Dirac points line up. In this way, contact between the low-energy bands at the moment of opening the gap happens along two parallel lines (Fig.~\ref{fig:1},~$b$). The value of the gap $\Delta$ is determined by the value of the interaction $U$ (Fig.~\ref{fig:1},~$c$).

%\newpage

\begin{figure}[!ht]
\centering
\vspace{6cm}
\includegraphics{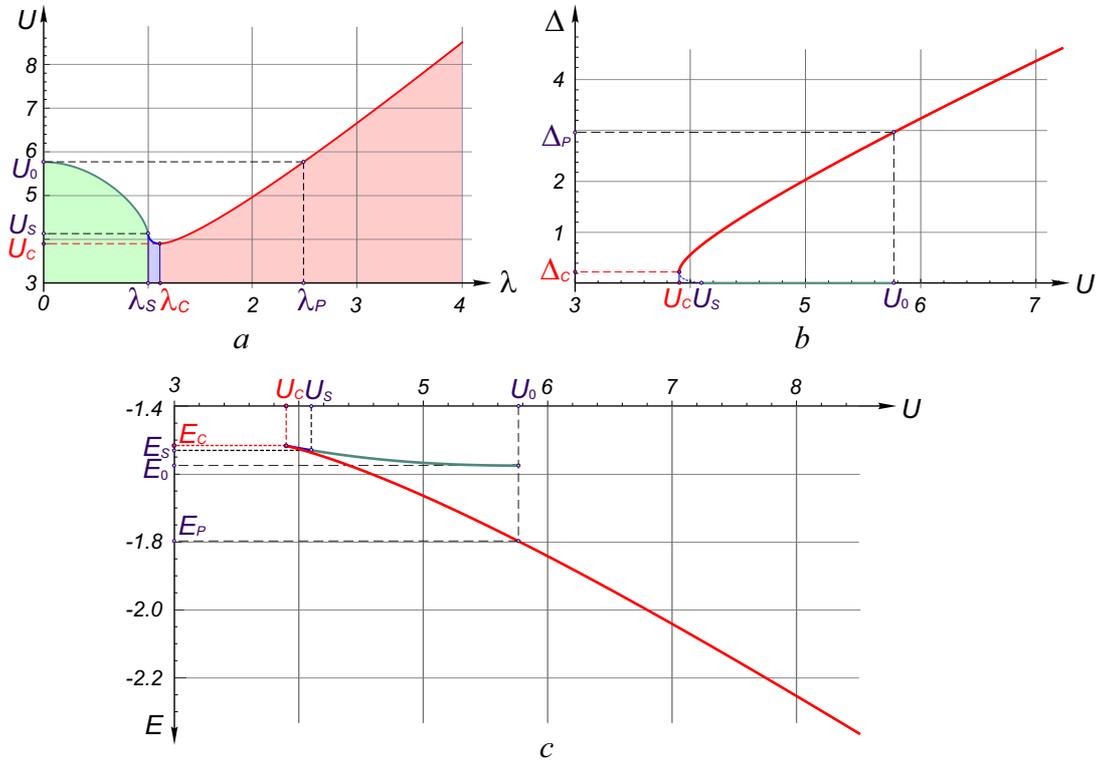}
%\vspace{-5cm}
\caption{(Color online) Numerical calculations of the on-site interaction $U$ as function of $\lambda$ $a$),
the gap $\Delta$ as function of $U$  $b$) and the density of the ground state energy $E$ as function of $U$ $c$). $U_0=5.767$, $E_0=-1.575$ ($\lambda =0$), $U_s=4.108$, $E_s=-1.530$, $\lambda_s=1$ (a gap opens), $U_c=3.904$ (the point of the Mott-Hubbard phase transition), $E_c=-1.517$, $\lambda_c=1.11$, $\Delta_c=0.22$, $E_p=-1.798$, $\lambda_p=2.484$, $\Delta_p=2.968$.  In Fig c) between $U_c$ and $U_s$ on the high energy curve the gapped state is unstable, between $U_s$ and $U_0$  on the high energy curve the gapless state is unstable, a  red curve corresponds to the stable state of the insulator. }
\label{fig:2}
\end{figure}

The behavior of electron liquid is not trivial near the point of the phase transition in insulator state. Equation (\ref{eq4}) connects the values of $\lambda$ and $U$, such $U$ has two solutions at $3.904<U<5.767$ and one solution at $U>5.767$ (see in Fig.~\ref{fig:2},~$a$). As we noted above the gap opens at $\lambda_s=1$, this point  corresponds to  $U_s=4.108$. When $\lambda$ varies from 1.110 to 0 first solution for $U$ varies in the interval $4.108<U<5.767$ is corresponds to a gapless state.  For $\lambda>1$ second solution describes an insulator state (see in Fig.~\ref{fig:2},~$a$). At $T=0$~K the stable phase state is determined by a minimal energy of electron liquid, its phase state  is determined by the ground state energy at differ $U$. The calculations of the density of the ground state energy are shown in Fig.~\ref{fig:2},~$c$: a high energy curve describes unstable state and a low energy curve defines a stable insulator state.  We have obtained, that the insulator state with $1<\lambda<1.11$ is unstable (see blue curve in Fig.~\ref{fig:2},~$c$). According to calculations a stable insulator state starts with a branch point at $U_c=3.904$ (see in Fig. 2, $a$). In this point a value of the gap in equal to $\Delta_c=0.22$ (see in Fig.~\ref{fig:2},~$b$),  it is a minimal value of the gap, in the $U$-infinity limit $\Delta \to  U$. As a rule, the Mott transition calculated on the lattices with simple symmetry is a second-order phase transition \cite{Arovas,Cyrot,Meng}, but in a hexagonal lattice the phase transition in the insulator state is the transition of the first-order at $U_c=3.904$.

%\newpage

We have shown that the nontrivial solution for wave vector $\textbf{q}$ corresponds to the minimum  of energy of the model (\ref{eq1}).
At half-filling the electron liquid on the hexagonal lattice with weak short-range repulsion is in a stable semi-metallic state. In the strong coupling limit a Mott insulator phase state is present. From numerical calculations of the Hubbard model which defined on the hexagonal lattice follow, that  the point of the phase transition corresponds to a critical value of the on-site Hubbard interaction $U_c$: $U_c=3.78$ \cite{Assaad}, $U_c=3.869 $ \cite{Sorella1}, $U_c = 4.5\pm 0.5$~\cite{Sorella2}, in the range 4–5 \cite{Paiva}, most likely the value of $U_c$ is close to 4.
The nature of the phase transition between these two states has not yet been established; the question about the order of the phase transition is open.  According to \cite{Meng} the transition between semi-metal and insulator phase states is realized via a gapped spin liquid  phase in an intermediate coupling regime $3.5<U<4.3$. In this case this transition is a first-order phase transition. Other numerical calculations (see for example \cite{Assaad,Sorella1,Sorella2,Paiva,Herbut}) show that we are dealing with the phase transition of the second-order.

\section{A triangular lattice}
\label{sec:3}

In this section we consider the behavior of electron liquid, which is described by the Hamiltonian (\ref{eq1}) defined on a triangular lattice. Using the same formalism we calculate the Mott-Hubbard phase transition in the model with the Hamiltonian (\ref{eq1}). The spectrum of the quasi-particle excitations has the following form

$$\varepsilon_{1,2}(\textbf{k,q})=-\frac{1}{2} \left(w(\textbf{k})+w(\textbf{k+q}) \pm\sqrt{[w(\textbf{k})-w(\textbf{k+q})]^2 + 4\lambda^2} \right),$$

where $w(\textbf{k+q}) = 2\cos(k_x+q_x) + 4\cos[(k_x+q_x)/2)]\cos[(k_y +q_y)\sqrt{3}/2]$, the wave vector $\textbf{q}$ has an nontrivial $q_x=0, q_y=2\pi /\sqrt{3}$ and $q_x=2 \pi, q_y=0$ values, that correspond the cell doubling.

\begin{figure}[!ht]
\centering
\includegraphics{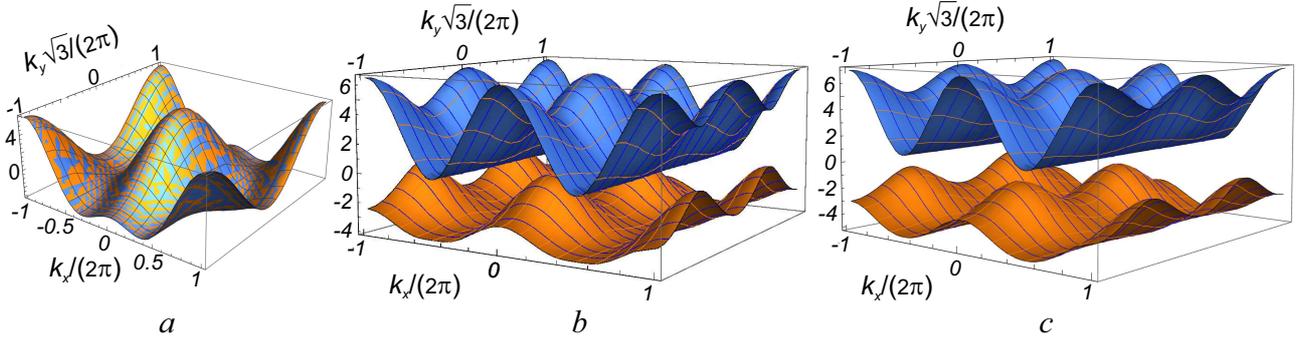}
%\vspace{0.5cm}
\caption{(Color online) Electron spectrum is calculated on the triangular lattice at $\lambda=0$ (gapless state) $a$), at $\lambda_c=2$ (state when the gap opens) $b$) and at $\lambda=2.5$ when gap is opened, here $\Delta=1$ $c$)}
\label{fig:3}
\end{figure}

We are investigated the phase transition in a half-filled electron liquid on a triangular lattice under the action of the on-site Hubbard repulsion $U$. The short-range repulsion potential leads to a qualitative change in the electronic spectrum. If $\lambda=0$ the energy spectrum has one branch (Fig.~\ref{fig:3},~$a$), which taking into account the on-site Hubbard repulsion is converted into a two branches, that is energy spectrum doubles (Fig.~\ref{fig:3},~$b$).  In the insulator state, these two branches of the spectrum are separated by the gap (Fig.~\ref{fig:3},~$c$). The spectrum remains asymmetric with respect to zero energy both in the gapless state (Fig.~\ref{fig:3},~$b$) and in the gapped state (Fig.~\ref{fig:3},~$c$).

At half filling, the minimum of the ground state energy of the system corresponds to a phase state with momenta $q_x=0, q_y=2\pi /\sqrt{3}$. This state, under the effect the effective $\lambda$-field leads to a doubling of the lattice cell (Fig.~\ref{fig:3},~$b$) and the formation of two branches of quasi particle excitations. For all $\lambda \geq 2$, the gap $\Delta$ opens $\Delta=2(\lambda-2)$, as a result of which the Mott transition in insulator phase state occurs.

%\newpage

\begin{figure}[!ht]
\centering
\vspace{5cm}
\includegraphics{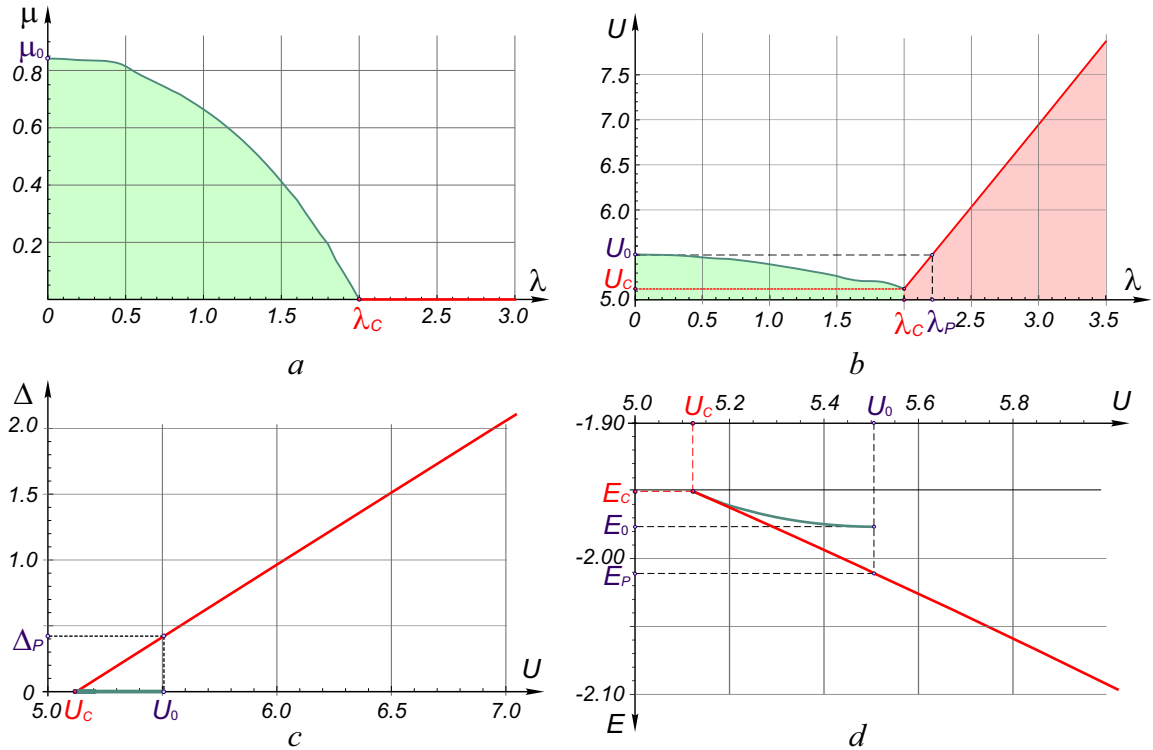}
%\vspace{-4.5cm}
\caption{(Color online) Numerical calculations of the chemical potential $\mu$ as function of $\lambda$ $a$), the on-site interaction $U$ as function of $\lambda$ $b$),
the gap $\Delta$ as function of $U$  $c$) and the density of the ground state energy $E$ as function of $U$ $d$). $U_0=5.507$, $E_0=-1.977$ ($\lambda=0$), $U_c=5.125$ (the point of the Mott-Hubbard phase transition at $\lambda_c=2$), $E_c=-1.951$, $E_p=-2.011$, $\lambda_p=2.211$, $\Delta_p=0.422$}
\label{fig:4}
\end{figure}

We accent our attention on the phase state of the electron liquid at half filling. The $q_x=0, q_y=2\pi /\sqrt{3}$ value does not break the chiral symmetry of considering phase states. The chemical potential monotonic decreases from maximum value $\mu_0=0.842$ at $\lambda=0$ to zero at $\lambda_c=2$, at $\lambda_c=2$ the gap in the spectrum opens and chemical potential is frozen on zero for arbitrary $\lambda\geq 2$ (see in Fig. ~\ref{fig:4},~$a$).

%\newpage

As in the hexagonal lattice the behavior of electron liquid is not trivial near the point of the phase transition ($\lambda_c=2$, $U_c=5.125$, $E_c=-1.951$). According to the Eq~(\ref{eq4}) the on-site Hubbard repulsion $U$ has two solutions at $5.125<U<5.507$ and one solution at $U>5.507$ (see in Fig.~\ref{fig:4},~$b$). But already at $\lambda=2$ the gap opens,  at  this point energy  equals to  $E_c=-1.951$ (Figs.~\ref{fig:4},~$c$ and  $d$). When $\lambda$ varies from 2 to 0, first solution for $U$ varies in the interval $5.125<U<5.507$ it corresponds to a gapless state.  For $\lambda \geq 2$ second solution describes a gapped state (see in Fig.~\ref{fig:4},~$b$). The stable phase state is determined by a minimal energy of electron liquid. The calculations of the density of the ground state energy are shown in Fig.~\ref{fig:4},~$d$: a high energy curve from $E_0=-1.977$ to $E_p=-2.011$ describes unstable state, a lower energy curve defines a stable insulator state.  According to calculations a stable insulator state starts with a branch point at $U_c=5.125$ (see in Fig. 4, $d$). In this point a value of the gap in equal to zero (see red curve in Figs.~\ref{fig:4},~$c$ and $d$).

Numerical calculations of the Mott-Hubbard phase transition from a paramagnetic metal state to an insulator state with the doubled cell show that it is a second-order phase transition. Critical value of the on-site interaction is equal $U_c =5.125$ (see in Fig.~\ref{fig:4},~$d$).
Thus, the Mott phase transition occurs in one scenario, regardless of the symmetry of the lattice;  a namely, the transition from a paramagnetic metal state (or semi-metal in the hexagonal lattice) to the insulator state, in which the unit cell doubles.
The ground state phase diagram of the Hubbard model defined on a triangular lattice does not taking into account the phase transition in the antiferromagnetic  gapped state in the calculations presented in \cite{Kokalj,Yoshioka}. Numerical calculations give the following values  of critical values of the on-site Hubbard repulsion $U_c = 7.5$ \cite{Kokalj}, $U_c = 7.4$ \cite{Yoshioka} at phase transition from a paramagnetic metal state to a nonmagnetic insulator state. Our calculations show that the phase transition from  a paramagnetic metal phase to the insulator phase occurs with doubled cell, as is the case for other lattice symmetries and dimension of the system \cite{Arovas,Cyrot,Meng,Sorella1,Paiva}.

\section{Conclusion}
\label{sec:4}

At a finite value of  the on-site interaction  $U>U_c$ and a half-filled occupation,  electrons with different momenta shifted by $\pi$ form the gapped spectrum of quasi-particle excitations. A stable insulator state is characterized by a doubled lattice cell what is an universal characteristic (for different dimension and symmetry of the lattice) of the insulator state.
The calculated values of $U_c$, $U_c=3.904$ for  hexagonal and $U_c=5.125$ for triangular lattices, are in good agreement with the numerical calculations of the Hubbard model on a hexagonal lattice.
We have shown that in a hexagonal lattice there is a first-order Mott-Hubbard phase transition from a semi-metallic phase state to an insulator state, while a second-order phase transition takes place for a triangular lattice.
At the same time, questions arise about the ground state phase diagram of the Hubbard model on a triangular lattice, since the insulator phase state with the doubled cell has not yet been numerical studied.
The proposed approach allows us to describe the Mott-Hubbard phase transition for the electron models with short-range repulsion for arbitrary dimension and symmetry of the lattice.\\

The research was partially supported by the programs of the National Academy of Sciences of Ukraine KPKVK 6541230-1A 'High-Strength States as a Result of Specific Interatomic Interaction in High-Entropy Solid Solutions with Martensitic Transformation' (State Reg. No.~0120U000160) and 6541230-3A 'Quantum Dynamics of Quasiparticle Excitations in Hybrid Metal Nanostructures' (State Reg. No.~0120U000132).

\end{document}